# Resisting hostility generated by terror: an agent-based study

Huet S. [1], Deffuant G. [1], Nugier A. [2], Streith M. [2], Guimond S. [2]

[1] LISC, Irstea, 9 av. Blaise Pascal, 63178 Aubière, France – sylvie.huet and
guillaume.deffuant@irstea.fr
[2] LAPSCO, UCA, 17 rue Paul Collomp, 63037 Clermont-Ferrand, France
armelle.nugier, michel.streith and serge.guimond@uca.fr

**Abstract.** We propose an agent-based model of cultural dynamics inspired from the social psychological theories. An agent has a cultural identity made of most acceptable positions about each of the different cultural worldviews corresponding to the main cultural groups of the considered society and a margin of acceptance around each of these most acceptable positions. An agent forms an attitude about another agent depending on the similarity between their cultural identities. When a terrorist attack is perpetrated in the name of an extreme cultural identity, the agents which are perceived negatively from this extreme cultural identity modify their margins of acceptance in order to differentiate themselves more from the threatening cultural identity. We initialized agents' cultural identities with data given by a survey on groups' attitudes among a large sample representative of the population of France, and then simulated these agents facing a threat. While in most cases, agents' attitude become more negative toward the agents belonging to the same group as the terrorists, we notice that when the population shows some cultural properties, the opposite takes place: the average attitude of the population for the cultural group to which the terrorists argued to belong becomes less negative. These properties are identified and explained. They especially permit to non-terrorists agents assimilated to terrorists to differentiate from them, and to other agents to perceive this change.

**Keywords:** Intergroup hostility, culture dynamics, Terror Management Theory, self-opinion

Cultural worldviews are defined as "shared conceptions of reality" [1, 2]. They are an important defense mechanism allowing people to cope with existential threats. This is why people are motivated to maintain faith in them. Terror Management Theory (TMT) has shown that a death fatality reminder, such as a terrorist attack, is a cultural threat or a self-worth threat [3]. Such a threat generally leads to an increase in negative intergroup bias in order to defend one's cultural worldviews [4, 5]. However, recent research showed that "increased prejudice and hostility are not an inevitable response to existential threat" [6]. Some cultural properties, when becoming salient simultaneously with the threat, increase perceived similarity of members of different groups, and protect against an increase of the intergroup hostility [6, 7]. Understanding when and why people react to a cultural or collective threat one way or the other is a basic problem having widespread theoretical and practical implications. To deal with this paradox, we study how the simulated change of cultural worldviews due to a cultural threat impacts virtual groups' attitudes toward each other. Our agent-based model helps to characterize cultural properties leading to acceptance or hostility between groups.

Agent-based model of culture dynamics have been seminally introduced by Axelrod [8]. The Axelrod model represents a culture as a set of traits and changes an unshared cultural trait by two agents to a shared one, with a probability depending on their level of shared properties. Several variants have been studied [9], introducing also a process leading traits to be more different [8, 10-13] instead of being shared. However, none of these models consider the impact of an existential threat on the cultural properties in relation with the self-worth and the group dynamics. This conclusion is also true for opinion dynamics model, very close from the cultural models. Though some of them include the possibility of rejecting the other's opinion instead conforming [12], or some rules for the evolution of the self-worth [14, 15], none of them address cultural properties, self-worth and group dynamics in relation to each other.

We model agents facing a cultural threat with dynamics of attitudes inspired by the general principles of the TMT. An agent has a segment of tolerance for each of the main cultural worldviews available in its environment and forms an attitude about the other agents depending on the similarity of acceptance segments about these worldviews. We assume that a terrorist attack is related to some extreme acceptance segments and related attitudes about the worldviews that are perceived as a threat by some agents, leading to differentiation from these extreme positions. These changes in acceptance segments modify the attitudes that the agents have about each other. We initialize the population of agents from aggregated data given by a representative survey on groups' attitudes conducted in France in 2014 [16] and then we submit it to a virtual threat. We generally observe an increase of hostility toward the group assimilated to the terrorist's group except in some particular cases where, on the contrary the hostility decreases. We study the evolution of the population in relation to its initial properties and propose some explanations to these variations.

The next section presents the model as well as the material. Then section 2 shows how the model is initialized and parameterized. Section 3 gives details about the evolutions and the related cultural properties. We finally conclude and discuss our results.

# 1 Method and materials

## 1.1 The model

### Hypothesis

This model is based on the idea of cultural worldview given by [3] and inspired by the Social Judgement Theory (SJT) [19]. We now propose an overview of the concepts that we use and their translation into the model.

- *Cultural worldview.* We assume that $K$ cultural worldviews are available in the environment. A cultural worldview is a consistent set of concepts, beliefs, tradi-tions or rituals organizing the world and agent behaviour. For instance we consider Christian, Muslim and areligious worldviews.

- *Agent position about a worldview.* Each agent has a most acceptable position about each worldview which is defined on a continuous axis from -1 (very negative), to +1 (very positive). This most acceptable position expresses her attitude about the considered worldview, i.e how much she likes/adheres, or dislikes/rejects it. It can be related to the most acceptable position of the SJT [19]. Note that an agent may have positive most acceptable positions about several worldviews. For instance, an agent can have a high positive most acceptable position for the areligious worldview, and also have a lower positive most acceptable position for Christian or Muslim worldviews.

- *Lower and higher margin of acceptance, acceptance segment of an agent for a worldview.* In addition to her position, an agent has margins of acceptance around it. The lower margin of acceptance is a segment of the [-1,1] axis which is lower than the agent most acceptable position, and the higher margin of acceptance is a segment defined similarly higher than the agent's most acceptable position. For convenience, the lower margin of acceptance is called margin(l) in the following, while the higher margin of acceptance is called margin(h). The segment defined by the union of the lower and higher margins of acceptance and the most acceptable position is called the acceptance segment. The positions located in this segment are assumed acceptable for the agent, while the positions located outside are not acceptable and perceived negatively.

- *Cultural identity of an agent.* The segments of acceptance for the *K* available worldviews represent the cultural identity of an agent [17, 18]. They indeed express the agent's own positions about the different worldviews and the positions that it considers acceptable.

- *Attitude of an agent about an acceptance segment for a worldview.* We compute an agent's attitude about an acceptance segment for a worldview applying a similarity function between this segment and her own acceptance segment for this worldview. The result is positive if the segments strongly overlap and more and more negative when they are separated and far apart. This is supported by Social judgement theory [19] which has shown "… the perceived distance depends on the level of involvement and the width of the latitude of acceptance." [21, 22].

- *Attitude of an agent about a cultural identity.* The attitude about a cultural identity is the average of the attitudes about the acceptance segments for the worldviews. To summarize, the more an agent perceives the cultural identity of another to be similar, i.e. in agreement to its own, the more its attitude about this agent is positive [6], whereas perceived differences lead to negativity. Indeed, "*people exaggerate the value of those who share their worldview or who provide positive evaluations and denigrate the value of those with diverging worldviews or who provide negative evaluations.*" [20].

- *Cultural group of an agent.* We assume that, if asked to declare her membership in a cultural group (such as Christian, Muslim, areligious), the agent would answer the one corresponding to the worldview for which his position is the highest.

We use this model to simulate how a terrorist attack may change the cultural identities of the agents and, as a consequence, the opinions attitudes of agents about each other (supposing that they are aware of all changes). We suppose that when perpetrating an attack, terrorists stress their cultural identity, for instance an extreme positive position for the Islamic worldview, extreme negative positions for the other worldviews and very narrow margins of acceptance around all these positions. We assume that the agents whose cultural identity is perceived negatively by the terrorists [17, 18], feel threat for their identity [3] and decrease their margins of acceptance in order to differentiate themselves from the terrorist cultural identity. Moreover, we assume that they reduce their margins more strongly when they perceive the terrorist identity relatively close to their own, (though far enough to be threatening) than when they are already very different.

### The cultural identities of agents

We consider a population of $N$ agents, each agent $i$ has a cultural identity defined by:

- $K$ values between -1 and +1, corresponding to its most acceptable position $a_K^i$ pro or anti on each of the $K$ cultural worldview present in the population. In the simulated populations, for each agent there is at least one worldview for which the agent has a positive most acceptable position. The identity group of an agent is defined by the worldview for which it has the highest most acceptable position.
- Moreover, the agent $i$ is also defined by margins of acceptance higher $B_K^i$, and lower $b_K^i$ its most acceptable positions which define a segment of acceptance indicating the acceptable positions for the agent. The segment going from $a_K^i$ to $B_K^i$ is called margin(h). The segment going from $b_K^i$ to $a_K^i$ is called margin(l). These segments are included in [-1;+1].

### Differences between cultural identities determine the attitude of agents about each other

Attitudes of agents about each other are computed from the comparison of their cultural identities. Each agent perceives its environment through its cultural identity and its attitude about another agent depends also on its perception of the other agent's cultural identity. Its attitude about its own cultural identity is at the maximum value: 1. An agent's attitude toward another agent's cultural identity is an average of the attitudes its gives to the perceived segments of acceptance composing the other agent's cultural identity. Indeed the agent perceives another agent's segments of acceptance as a discretized segment of positions $a$.

Then, considering a worldview $K$ of an agent $i$, its related most acceptable position and margins of acceptance allow it to compute the attitude $\omega_i^K(a)$ about a position $a$ of the worldview $K$ as follows:

a) $\quad \omega_K^i(a) = 1, \text{ if } a < B_K^i \text{ and } a > b_K^i$ \hfill (1)

b)  $\omega_K^i(a) = \frac{e^y - 1}{e^y + 1}$ with  $y = 1 + \frac{a - a_K^i}{a_K^i - b_K^i}$, if $a \leq b_K^i$

c)  $\omega_K^i(a) = \frac{e^y - 1}{e^y + 1}$ with  $y = 1 + \frac{a_K^i - a}{B_K^i - a_K^i}$, if $a \geq B_K^i$

We suppose that $\omega_K^i(a)$ equals +1 if $a$ is in the margins of acceptance of $i$. If not, $\omega_K^i()$ is 0 for $a$ at the bounds of the segment of acceptance and it decreases with the distance between $a$ and its closest bound of the segment of acceptance, with an asymptote at -1. Moreover, the decrease to -1 is faster for smaller margin of acceptance. When the decrease is fast, the opinion of the agent can be almost the same (close to -1) for different values of attitude $a$. This function behavior is illustrated in figure 1.

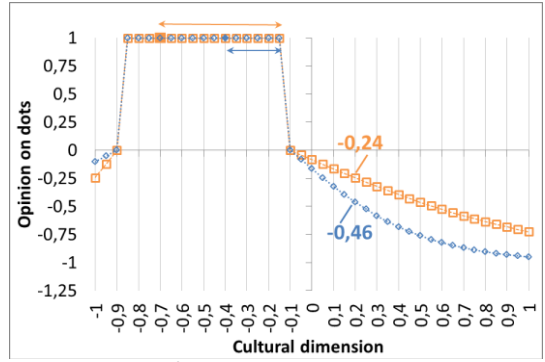

**Fig. 1.** Computation of the attitude $\omega_K^i(a)$ related to each position $a$ (in abscissa) for two agents with a similar segment of acceptance valued 1 for every $a$ of their segment, but a different width of margin(h) (see the orange arrow for the orange agent, and the blue arrow for the blue agent). We see the agent having the smallest width, the blue one, going more quickly to highly negative values than the orange agent for every $a$ outside their segment

The axis of each worldview $K$ is divided regularly into $d$ values $a_p$(from -1 to +1) and agent $i$ computes its attitude about the $K$ acceptance segment of agent $j$, its average attitude for the positions $a_p$ located in the acceptance segment of the agent $j$. More formally, the opinion of $i$ about $j$ 's acceptance segment in worldview $K$ is given by:

$$\omega_K^{ij} = 2 \frac{\sum_{p=1}^{d} \omega_K^i(a_p) \max(\omega_K^j(a_p), 0)}{\sum_{p=1}^{d} \max(\omega_K^j(a_p), 0)} \qquad (2)$$

Note, that $\omega_K^{ij}$ is maximal when the two acceptance segments are identical.

Finally the overall attitude of $i$ about $j$ is the average attitude over the #$K$ different acceptance segments designing totally the cultural identity of an agent:

$$\omega^{ij} = \frac{\sum_{K=1}^{\#K} \omega_K^{ij}}{\#K} \qquad (3)$$

Qualitatively, the agents with large margins of acceptance tend to have a positive attitude of most of the others, whereas agents with small margins of acceptance are very selective and have very negative attitudes about many others.

**Impact of a threat on agent's cultural identity**

A terrorist attack is modeled by a scenario of messages in the media, conveying the cultural identity of terrorists. The most acceptable positions of terrorists are assumed to be very positive for one worldview and very negative about the others. Moreover their margins of acceptance are assumed to be very small.

The agents of the population perceiving as negative the attitude of terrorist about their cultural identity are assumed to perceive a "threat" and modify their margins of acceptance. More precisely, let the most acceptable positions of terrorist be defined by values $(a_K^q, b_K^q, B_K^q)$, the agents compute $\omega^{qi}$ $q$ the attitude of terrorist $q$, about them. If this attitude is positive, $i$ is not scared and does not react. But if $\omega^{qi}$ is negative, $i$ modifies its margins of acceptance away from the acceptance segments of $q$ as follows.

If $\omega^{qi} < 0$, the intensity of the margin of acceptance modification $\mu$ is:

$$\mu = \alpha \frac{e^{\omega^{qi}} - 1}{e^{\omega^{qi}} + 1} \qquad (4)$$

where $\alpha$ is a positive number smaller than 1. The value of $\mu$ is close to -1 when $\omega^{qi}$ is very negative.

For $K$ equal to the main acceptance segment of the terrorist, i.e. the aggressor, the bound $b_K^i \in \{b_K^i, B_K^i\}$ which is the closest to $a_K^q$ is modified as follows (with $t$ equals to the time):

$$b_K^i(t+1) = b_K^i(t) + \mu(b_K^i(t) - a_K^i - \varepsilon \frac{|b_K^i(t) - a_K^i|}{b_K^i(t) - a_K^i})) \qquad (5)$$

where $\varepsilon$ is a small positive number representing the smallest possible margin of acceptance width (parameter of the model).

This change in the margins of acceptance results in a more negative attitude about the aggressor, in accordance with experimental observations [20].

## 1.2    Data source

A survey on group's attitudes in France has been used to initialize the model [16]. We wanted our virtual groups composing our populations to respect the measured ingroup and intergroup attitudes. The surveyed sample includes 1000 people, representative of the French population. People answered to the question « what is your general attitude about the following groups? ». The groups are A (for Areligious), M (for Muslims), and C (for Christians). They have to answer using a five-point scale ranging from "strongly unfavorable" to "strongly favorable". Averages and standard-deviations for in-group attitudes and every out-group attitudes have been computed and normalized between -1 to +1 for modeling purpose. Figure 2 shows the groups' attitudes.

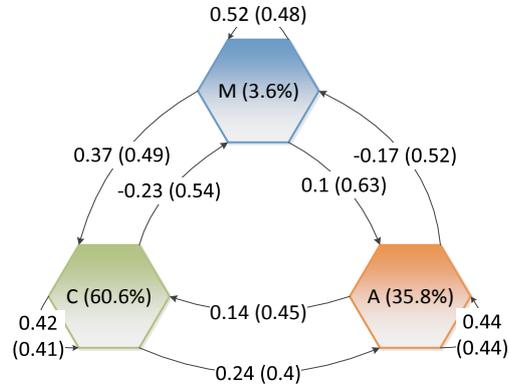

**Fig. 2.** Attitudes of each group M (for Muslims), C (for Christians), A (for Areligious) toward each other (averages and standard-deviations in parenthesis for each relation computed from the representative survey of the French population)

## 2 Initialisation of the virtual population and parametrization of the model

### 2.1 Building virtual initial populations

Our population of virtual agents, have a cultural identity defined by their acceptance segment for 3 cultural worldviews $C$, $M$ and $A$. The worldview for which the agent has the highest most acceptable position defines the agent's cultural group.

We assume that each group includes agents of two types:

- Agents with exclusive identities: these agents have their acceptance segment in the positive side for the worldview $K \in \{M, C, A\}$ defining their group, and on the negative side for the other worldviews. Amongst these agents, we identify the extremists which have most acceptable positions close to 1 or -1 and narrow acceptance segments.
- Agents with inclusive identities: these agents have one or two most acceptable which are positive and the corresponding margin of acceptance in the positive side and the one or two others close to zero, with a margin of acceptance also mostly in the positive side

We assume that each group ($C$, $M$ or $A$) of the initial population includes $x\%$ of inclusive and $y\%$ of exclusive agents having their highest most acceptable position for the worldview defining their group. The values $x$ and $y$ are determined to fit indicators extracted from the survey.

Thus the set of values defining the 3 acceptance segments of each one of the 6 "prototypical" agents, as well as every $x$ and $y$ associated to each group are determined by an optimization minimizing the distance between indicators computed from the virtual population and references. Indeed, to build a virtual initial population close enough to the aggregated indicators (averages and standards deviations) computed from the data of the survey, we minimize the sum of the absolute distances to the references of the corresponding indicators measured for the virtual populations.

From the optimization results, we keep the 120 best virtual populations having relative average errors (over all the distance to the references) going from 5% to 7%, and maximum errors going from 21 % to 44 %.

## 2.2     Parameterizing the evolution of the population.

Once we have virtual populations, we want to simulate their evolution facing a set of "threat" messages conveying the cultural identity of a terrorist.

The tested scenario comprises seven consecutive similar terrorists' messages. The terrorists have very narrow acceptance segments with a very positive most acceptable position for the worldview $M$ defining their group and very negative most acceptable positions about the two other worldviews ($C$ and $A$). Thus they tend to have a very negative attitude about all other cultural identities.

The following section presents how the agents change their margins of acceptance when they receive the messages. The parameters of the dynamics take the following value: $\alpha$=0.5; $\varepsilon$=0.05 and $d$ = 400.

## 3      Evolution of virtual group attitudes facing "threats"

We simulate the evolution of our 120 populations facing a set of seven consecutive "threat" messages. At each step, due to the high level of mediatization and related discussions after a terrorist act, agents are considered as perfectly informed about the other's current acceptance segments. We investigate the result to identify how the average attitude of the population about the cultural identities of the inclusive and exclusive agents of group $M$, which is also the group of the terrorists, evolves.

They show that in most of the selected populations, the average attitude about group $M$ cultural identities decreases, while for some of them it increases on the contrary. The attitudes about the other groups $C$ and $A$ remain rather stable on average. The challenge is thus to understand why and when a population becomes more positive toward $M$ after facing a threat coming from agents from this group.

From the analysis of the results shown in figure 3, we obtain more information about which agent decreases or increases its attitude about group $M$. We see at first that the majority of each group decreases its attitude toward inclusive and exclusive $M$ (in white). Complementary, we see also these are mainly the agents of group $C$ which generally become more positive, about inclusive $M$ (in red) only, but also sometimes about exclusive $M$ (if $C$ are inclusives) (in dark blue). Moreover, we observe that agents of group $A$ can also sometimes increase their attitude about inclusive $M$ (in red), but never about exclusive $M$. We observe also a minority of $M$ agents increases their attitude about exclusive $M$ and decrease it for inclusive $M$ (in light blue).

We then focus on agent's cultural properties explaining these observations. During the simulations, agents change their higher margin of acceptance in their attitudinal segment on the *M* worldview. In populations increasing their average attitude about group *M*, we notice that:

- *M* agents decrease strongly their higher margin of acceptance on the *M* worldview;
- Non-*M* (especially *C*) agents decrease moderately their higher margin of acceptance on the *M* worldview.

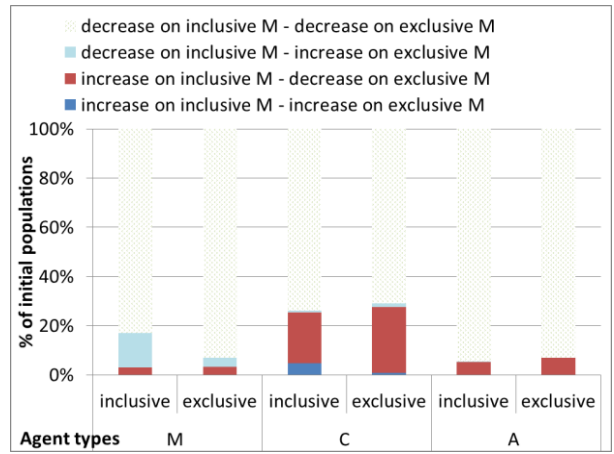

**Fig. 3.** Distribution in percentage of all the agents' types of change of attitude over all the 120 initial populations and times

Going further in the observation of data, we establish that the following set of conditions is met altogether in these populations at the time 0 and/or later due to the impact of the threat:

- ***For the M agents***: the attitude toward *M* worldview tends to be lower and _ ***the margin(h – for higher)) of acceptance larger -*** than in the average population. Moreover, a ***large margin(l - for lower) of acceptance*** is necessary to get an increase of attitude from others. Indeed, these properties lead to decrease strongly the margin(h) of acceptance when facing an *M* extremist cultural identity. The margin(l) of acceptance has to be large enough to ensure an overlap of its acceptance segment with the ones of non-*M* agents, overlap which is necessary for a positive attitude.
- ***For the non-M agents***: the attitude about *M* worldview is higher, the margin(h) of acceptance is smaller and the margin(l) of acceptance is larger than in the average populations. ***The small margin(h) of acceptance (condition 1) indeed ensures that this margin is weakly modified by the threat***. ***Their high attitude and their large margin(l) of acceptance (condition 2) ensure*** an overlap of their acceptance segment with the one of the group *M* agents (mainly inclusive).

The next figures illustrate why these conditions lead to increase the attitude about the group with the same main worldview as the terrorists. We consider a standard case when the expected outcome of decreasing average opinion about group $M$ takes place. In the following figures, the attitudinal function on the $M$ worldview of a group $A$ or $C$ exclusive or inclusive is represented in orange while the attitudinal function of a group $M$ inclusive or exclusive agent is represented in blue. The left panel represents the attitudinal funtion before the aggression and the right panel this function after the aggression.

Figures 4 illustrate what occur when there isn't a small margin(h) of acceptance of non-$M$ blue agent (this is generally a property of inclusive agents). We clearly see on the figures that despite the increasing part of the overlap in the inclusive $M$ acceptance segment, the decreasing of the margin(h) of acceptance of the inclusive $A$ is such as it perceives the inclusive $M$ as very much further than before.

Figures 5 illustrate what occur when there isn't a large margin(h) of acceptance for blue $M$ agents (this is generally a property of exclusive agents). We observe that the attitude of the inclusive $C$ for the exclusive $M$ is not changed by the change of $M$ since the average view of $C$ for $M$ remains -1.

Figures 6 illustrate what occur when the condition 2 is not met. We have then a large margin(h) of acceptance for both $M$ and non-$M$ agents. Then, when the two agents ($C$ and $M$) reject the extremist's view, the non-$M$ agent becomes less tolerant (from 0.29 to -0.01 on the graphs)!

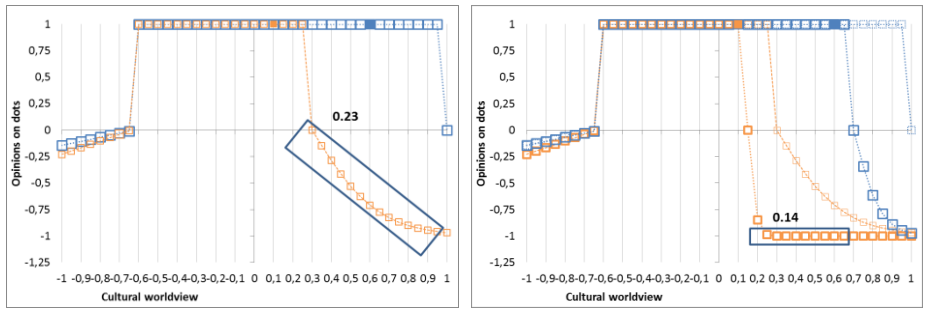

**Fig. 4.** Evolution of the attitude of inclusive $A$ (in orange) for inclusive $M$ (in blue): on the left before the terrorist attack, on the right, after the agression: the attitude decreases (from 0.23 to 0.14) due to the decreasing of the margin(h) of acceptance of the inclusive A and despite the part of the overlap over the inclusive $M$ acceptance segment has increased. Indeed, being less inclusive(h), it sees the part of the inclusive $M$'s segment external to it strongly more negatively than before.

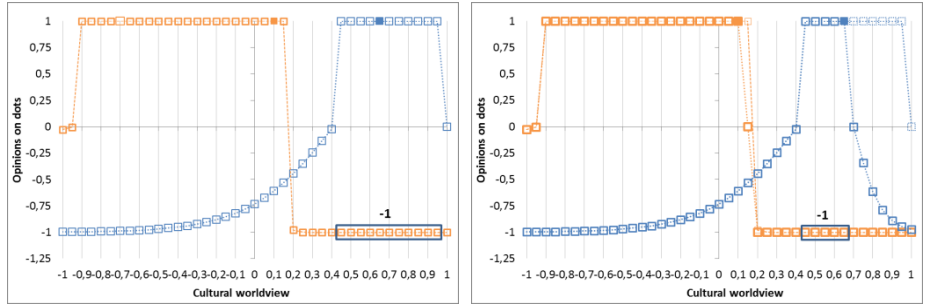

**Fig. 5.** Evolution of the attitude of exclusive *C* (in orange) for exclusive *M* (in blue): on the left before the aggression, on the right, after the agression: the attitude does not change despite the change of *M* margin(h) of acceptance

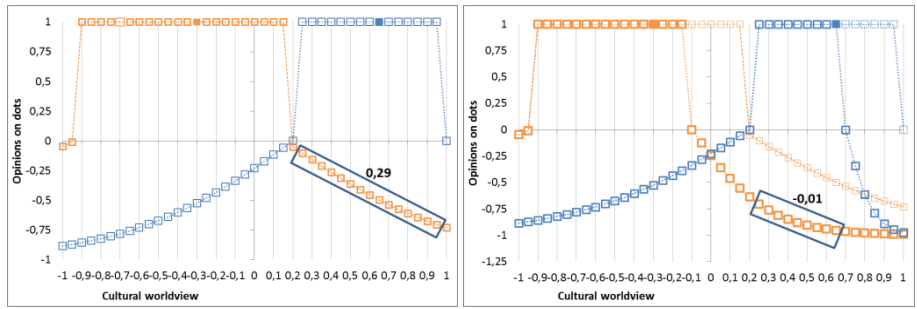

**Fig. 6.** Evolution of the attitude of exclusive *C* (in orange) for exclusive *M* (in blue): on the left before the attack, on the right, after the agression: most of the time, when the two agents (*C* and *M*) reject the extremist's view, the non-*M* agent becomes less tolerant (from 0.29 to -0.01 on the graphs).

## 4 Conclusion - Discussion

The modeled populations increasing their attitude about the group of main worldview *M* to which a terrorist group is assimilated, show specific initial acceptance segments about this worldview: (1) all agents tend to have a large margin(l) of acceptance; (2) group *C* and group *A* agents have a small margin(h) of acceptance and group *M* agents have a large margin(h) of acceptance; (3) group *C* and *A* agents have a high attitude about worldview *M* .

In other words, an average increase of attitude about the group of the terrorists takes place if:

- this group is mainly composed from very inclusive agents with a moderately positive attitude for their own group worldview. They also should have a large margin of acceptance for critical attitudes about their group worldview.
- the members of the other groups have a relatively high positive attitude about the terrorist group worldview, and their view is not affected by the attack. Moreover, they are tolerant, inclusive for attitudes different from theirs about the terrorist group worldview.

The large margin of acceptance for mean or mild attitudes about the terrorist group worldview increases the overlap between acceptance segments of the agents of the population. Such overlaps can be related to the feeling of wide similarity and solidarity observed in the large French demonstrations that took place after the Charlie Hebdo terrorist attacks. These conclusions are in line with the results of [6, 7]. Indeed, we observe that some cultural properties, when becoming salient simultaneously with the threat, may increase perceived similarity of members of different groups, and protect against an increase of the intergroup hostility.

**It should be underlined that, in the model, this solidarity takes place because the agents of the same group as the terrorists reject strongly the radical attitude of the terrorists, and that the other agents are aware of their evolution.** The next step of our work is to come back to experiment to check the relevance of the model behaviour.

Another step regarding the study of the model is a larger study of the sensitivity of the result to the distribution of inclusive versus exclusive agents in each group. Indeed, we have seen inclusive agents are more enclined to become less negative to *M* than the other agents. Then the proportion of such type of agent in the population is very important to observe an increase of the attitude. A variation of this proportion can probably change drastically the result.

## Acknowledgements

The authors acknowledge funding from the "Mission for Interdisciplinarity" of the CNRS.